\documentclass[12pt,a4paper]{article}
\usepackage{a4wide}
\usepackage{graphicx}
\usepackage{cite}
\begin{document}

\begin{center}
{\bf Microelectronics Reliability {\bf 40 } (2000) 1787--1790}
\end{center}

\vskip 2cm

\begin{center}
{\large {\bf On the intrinsic origin of 1/f noise }}
\\[1.25\baselineskip] {\bf B.\thinspace Kaulakys } \\
{\small Institute of Theoretical Physics and Astronomy, Vilnius University,\\ 
A.\thinspace Go\v stauto 12, LT-2600 Vilnius, Lithuania and } 
\\[0.25\baselineskip] {\small Faculty of Physics, Vilnius University, 
Saul\.etekio al.\ 9, LT-2040 Vilnius, Lithuania }
\end{center}

\vskip 0.2cm

\begin{abstract}

The problem of the intrinsic origin of $1/f$ noise is considered. Currents
and signals consisting of a sequence of pulses are analysed. It is shown
that intrinsic origin of $1/f$ noise is a random walk of the average time
between subsequent pulses of the pulse sequence, or interevent time. This
results in the long-memory process for the pulse occurrence time and in $1/f$
type power spectrum of the signal.
\end{abstract}

\vskip 1cm

\section{Introduction}

Ubiquitous of signals and processes with $1/f$ power spectral density at low
frequencies has led to speculations that there might exist some generic
mechanism underlying production of $1/f$ noise. The generic origins of two
popular noises: white noise (no correlation in time, $S(f)\infty 1/f^0$) and
Brownian noise (no correlation between increments, $S(f)\infty 1/f^2$) are
very well known. It should be noted, that Brownian motion is the integral of
white noise and that operation of integration of the signal increases the
exponent by 2 while the inverse operation of differentiation decreases it by
2. Therefore, $1/f$ noise can not be obtained by simple procedure of
integration or differentiation of the convenient signals. There are no
simple, even stochastic, equations generating signals with $1/f$ noise. Note
in this context also to the concept of the fractional Brownian motion and to
the half-integration of a white noise signals used for generation of
processes with $1/f$ noise \cite{Man68}. These and similar mathematical
algorithms, procedures and models for generation of the processes with $1/f$
noise \cite{Hig90,Nin98} are, however, sufficiently specific, formal or
unphysical. They can not, as a rule, be solved analytically and they do not
reveal the origin as well as the necessary and sufficient conditions for
appearance of $1/f$ type fluctuations. Physical models of $1/f$ noise in
some physical systems are usually very specialized, complicated and they do
not explain the internal origin of the omnipresent processes with $
1/f^\delta $ spectrum.

A lot of contributions are available in the literature concerning the origin
of $1/f$ noise. On the web (http://linkage.rockefeller.edu/wli/1fnoise),
Wentain Li is collective bibliography of flicker noise. Sufficiently
comprehensive bibliography of the contributions concerning the modeling of $
1/f$ noise may be find in references [3--7].

This work is a continuation of series of papers devoted to the modeling $1/f$
noise in simple systems \cite{KV95C,KM97C} and search of necessary
conditions for appearance of the signals with power spectrum at low
frequencies like $S(f)\infty 1/f^\delta $ ($\delta \simeq 1$) [6--9]. In
papers \cite{KV95C,KM97C} an analysis of the necessary conditions for
appearance of $1/f$ type fluctuations in the simple systems consisting of
few or even one particle and affected by random perturbations is presented.
Later, a simple analytically solvable model of $1/f$ noise has been proposed
\cite{K98}, analyzed \cite{KM98} and generalized \cite{KM98VC}. The model
reveals main features, and parameter dependencies of the power spectrum of $
1/f$ noise.

Here considering signals and currents as consisting of pulses
generalizations and development of the model \cite{K98} are presented. The
paper includes derivation of the expression for the correlation function,
analysis of the examples of different signals and exhibition of the
necessary conditions for appearance of $1/f$ type power spectrum in the
signals consisting of pulses. It is shown that intrinsic origin of $1/f$
noise is a Brownian motion of the pulse interevent time.

\section{The model}

Let us consider currents or signals represented as sequences of random (but
correlated) pulses $A_k(t-t_k)$. Function $A_k(t-t_k)$ represents the shape
of the $k$-pulse of the signal in the region of the pulse occurrence time $
t_k$. The signal or intensity of the current of particles in some space
cross section may, therefore, be expressed as
$$
I\left( t\right) =\sum\limits_kA_k\left( t-t_k\right) .\eqno{(1)}
$$
It is easy to show that the shapes of the pulses mainly influence the high
frequency, $f\geq \Delta t_p$ with $\Delta t_p$ being the characteristic
pulse length, power spectrum while fluctuations of the pulse amplitudes
result, as a rule, in the white or Lorentzian but not $1/f$ noise \cite
{Schick74}. Therefore, we restrict our analysis to the noise due to the
correlations between the pulse occurrence times $t_k$. In such an approach
we can replace the function $A_k\left( t-t_k\right) $ by $a\delta \left(
t-t_k\right) $. Here $\delta \left( t-t_k\right) $ is the Dirac delta
function, $a=\left\langle \int_{-\infty }^{+\infty }A_k\left( t-t_k\right)
{\rm d}t\right\rangle $ is the average area of the pulse and the brackets $
\left\langle \ldots \right\rangle $ denote the averaging over realizations
of the process. In such an approach the signal (1) my be expressed as
$$
I\left( t\right) =a\sum\limits_k\delta \left( t-t_k\right) .\eqno{(2)}
$$
This model also corresponds to the flow of identical point objects:
electrons, photons, cars and so on. On the other hand, fluctuations of the
amplitudes $A_k$ may result in the additional noise but can not reduce $1/f$
noise we are looking for.

\subsection{Power spectrum}

Power spectral density of the signal $I\left( t\right) $\ is
$$
S\left( f\right) =\lim \limits_{T\rightarrow \infty }\left\langle \frac
2T\left| \int\limits_{t_i}^{t_f}I\left( t\right) e^{-i2\pi ft}{\rm d}
t\right| ^2\right\rangle \eqno{(3)}
$$
where $T=t_f-t_i$ is the observation time.

We can also introduce the autocorrelation function
$$
\Phi \left( s\right) =\left\langle \frac 1T\int\limits_{t_i}^{t_f-s}I\left(
t\right) I\left( t+s\right) {\rm d}t\right\rangle \eqno{(4)}
$$
and use the Wiener-Khintchine relations:
$$
S\left( f\right) =4\lim \limits_{T\rightarrow \infty }\int\limits_0^T\Phi
\left( s\right) \cos \left( 2\pi fs\right) {\rm d}s,\eqno{(5)}
$$
$$
\Phi \left( s\right) =\int\limits_0^\infty S\left( f\right) \cos \left( 2\pi
fs\right) {\rm d}f.
$$

Substitution of Eq. (2) into Eq. (3) results the power spectral density of
the signal expressed as sequence of pulses
$$
S\left( f\right) =\lim \limits_{T\rightarrow \infty }\left\langle \frac{2a^2}
T\left| \sum_{k=k_{\min }}^{k_{\max }}e^{-i2\pi ft_k}\right| ^2\right\rangle
$$
$$
=\lim \limits_{T\rightarrow \infty }\left\langle \frac{2a^2}T\sum_{k=k_{\min
}}^{k_{\max }}\sum_{q=k_{\min }-k}^{k_{\max }-k}e^{i2\pi f\Delta \left(
k;q\right) }\right\rangle \eqno{(6)}
$$
where $\Delta \left( k;q\right) \equiv t_{k+q}-t_k$ is the difference of
pulse occurrence times $t_{k+q}$ and $t_k$ while $k_{\min }$ and $k_{\max }$
are minimal and maximal values of index $k$ in the interval of observation $
T=t_f-t_i$.

For a stationary process Eq. (6) yields
$$
S\left( f\right) =\frac{2a^2}T\sum_{q=-N}^N\left( N+1-\left| q\right|
\right) \left\langle e^{i2\pi f\Delta \left( q\right) }\right\rangle .
$$
Here $N=k_{\max }-k_{\min }$, the brackets $\left\langle \ldots
\right\rangle $ denote the averaging over realizations of the process and
over the time (index $k$) and a definition $\Delta \left( q\right) =-\Delta
\left( -q\right) =\Delta \left( k;q\right) $ is introduced. For abbreviation
of the equations we have omitted the mark of the limit ''$\lim
\limits_{T\rightarrow \infty }$'' here and further on in expressions for the
power spectrum $S\left( f\right) $.

The average over the distribution of $\Delta \left( q\right) $ may be
expressed as
$$
\left\langle e^{i2\pi f\Delta \left( q\right) }\right\rangle \equiv
\int\limits_{-\infty }^{+\infty }e^{i2\pi f\Delta \left( q\right) }\Psi
_\Delta \left( \Delta \left( q\right) \right) {\rm d}\Delta \left( q\right)
$$
$$
=\chi _{\Delta \left( q\right) }\left( 2\pi f\right) .
$$
Here $\Psi _\Delta \left( \Delta \left( q\right) \right) $ is the
distribution density of $\Delta \left( q\right) $, and $\chi _{\Delta \left(
q\right) }\left( 2\pi f\right) $ is the characteristic function of the
distribution $\Psi _\Delta \left( \Delta \left( q\right) \right) $.
Therefore
$$
S\left( f\right) =2a^2\sum\limits_{q=-N}^N\left( \overline{\nu }-\frac{
\left| q\right| }T\right) \chi _{\Delta \left( q\right) }\left( 2\pi
f\right) \eqno{(7)}
$$
where $\overline{\nu }=$$\left\langle \lim \limits_{T\rightarrow \infty }
\frac{N+1}T\right\rangle $\ is the mean number of pulses per unit time.

When the sum $\sum\limits_{q=-N}^N\left| q\right| \chi _{\Delta \left(
q\right) }\left( 2\pi f\right) $ converges and $T\rightarrow \infty $ we
have the power spectrum from Eq. (7) in the form
$$
S\left( f\right) =2a\overline{I}\sum\limits_{q=-N}^N\chi _{\Delta \left(
q\right) }\left( 2\pi f\right) .\eqno{(8)}
$$
Here $\overline{I}\equiv \left\langle \overline{I\left( t\right) }
\right\rangle =\overline{\nu }a$ is the average current.

\subsection{Correlation function}

Substitution of Eq. (2) into Eq. (4) yields the correlation function of the
signal (2)
$$
\Phi \left( s\right) =\left\langle \frac{a^2}T\sum\limits_{k,q}\delta \left(
t_{k+q}-t_k-s\right) \right\rangle .
$$
After summation over index $k$ we have
$$
\Phi \left( s\right) =\bar Ia\sum_q\left\langle \delta \left( \Delta \left(
q\right) -s\right) \right\rangle
$$
where the brackets $\left\langle \ldots \right\rangle $ denote again the
averaging over realizations of the process and over the time (index $k$), as
well. Such averaging coincides with the averaging over the distribution of
the time difference $\Delta \left( q\right) $
$$
\Phi \left( s\right) =\bar Ia\sum_q\int\limits_{-\infty }^{+\infty }\psi
_\Delta (\Delta )\delta \left( \Delta -s\right) {\rm d}\Delta
$$
$$
\Phi \left( s\right) =\bar Ia\delta \left( s\right) +\bar Ia\sum_{q\neq
0}\psi _\Delta (s).\eqno{(9)}
$$
Here $\psi _\Delta (\Delta )$ is the distribution density of $\Delta \left(
q\right) $. Substitution of Eq. (9) into Eq. (5) yields expressions (7) and
(8).

\section{Examples}

Consider some examples of the signals represented by Eq. (2).

{\bf (i)} Periodic signal expressed as $I\left( t\right) =a\sum_k\delta
\left( t-k\tau \right) $ generates the power spectrum
$$
\begin{array}{c}
S\left( f\right) =2a^2\lim \limits_{T\rightarrow \infty }
\frac{\sin ^2\left( \pi \left( N+1\right) \tau f\right) }{T\sin ^2\left( \pi
\tau f\right) } \Rightarrow 2\overline{I}^2\delta \left( f\right) ,\quad
f\ll \tau ^{-1}.
\end{array}
$$

{\bf (ii)}\ Perturbed periodic signal represented by Eq. (2) with the times
series expressed as recurrence equations $t_k-t_{k-1}\equiv \tau _k=
\overline{\tau }+\sigma \varepsilon _k$ with $\left\{ \varepsilon _k\right\}
$ being a sequence of uncorrelated normally distributed random variables
with zero expectation and unit variance and $\sigma $ being the standard
deviation of this white noise \cite{KM97C}. For this model we have $\Delta
\left( q\right) =q\overline{\tau }+\sigma
\sum\limits_{l=k+1}^{k+q}\varepsilon _l$\ and the characteristic function
$$
\chi _{\Delta \left( q\right) }\left( 2\pi f\right) =e^{{\rm i}2\pi
f\left\langle \Delta \left( q\right) \right\rangle -\frac 12\left( 2\pi
f\right) ^2\sigma _\Delta ^2}\eqno{(10)}
$$
where $\left\langle \Delta \left( q\right) \right\rangle =q\overline{\tau }$
and the variance $\sigma _\Delta ^2$ of the time difference $\Delta (q)$
equals $\sigma _\Delta ^2\equiv \left\langle \Delta \left( q\right)
^2\right\rangle -\left\langle \Delta \left( q\right) \right\rangle ^2=\sigma
^2\left| q\right| $. Substitution of Eq. (10) into Eq. (8) yields the
Lorentzian spectrum
$$
S\left( f\right) =\overline{I}^2\frac{4\tau _{{\rm rel}}}{1+\tau _{{\rm rel}
}^2\omega ^2}\eqno{(11)}
$$
where $\omega =2\pi f$ and $\tau _{{\rm rel}}=\sigma ^2/2\overline{\tau }$.

{\bf (iii)} Time difference $\Delta \left( q\right)
=\sum\limits_{l=k+1}^{k+q}\tau _l$ as a sum of uncorrelated interevent times
$\tau _l$. According to Eqs. (6)--(8) we have in this case
$$
S\left( f\right) =2a\overline{I}\left[ 1+2Re\sum_{q=1}^N\left\langle
e^{i2\pi f\tau }\right\rangle ^q\right]
$$
$$
S\left( f\right) =2a\overline{I}\left[ 1+2Re\frac{\chi _\tau \left( \omega
\right) }{1-\chi _\tau \left( \omega \right) }\right] .\eqno{(12)}
$$
For instance, substitution at $f\ll \overline{\tau }^{-1}$ and $f\ll \sigma
^{-1}$ of Eq. (10) with $q=1$ into Eq. (12) results in Eq. (11).

{\bf (iv)} For the Poisson process
$$
\chi _\tau \left( 2\pi f\right) =\frac 1{1-{\rm i2}\pi f\overline{\tau }
},\quad Re\frac{\chi _\tau \left( \omega \right) }{1-\chi _\tau \left(
\omega \right) }=0
$$
and we have from Eq. (12) only the shot noise
$$
S\left( f\right) =2a\overline{I}=S_{{\rm shot}}.\eqno{(13)}
$$

{\bf (v)} Brownian motion of the interevent time $\tau _k$ with some
restrictions, e.g., with the relaxation to the average value $\overline{\tau
}$,
$$
\tau _k=\tau _{k-1}-\gamma \left( \tau _{k-1}-\bar \tau \right) +\sigma
\varepsilon _k,\eqno{(14)}
$$
when the pulse occurrence times $t_k$ are expressed as
$$
t_k=t_{k-1}+\tau _k.\eqno{(15)}
$$
According to Eq. (6) the power spectrum of the signal (2) with the pulse
occurrence times $t_k$ generated by Eqs. (14) and (15) for sufficiently
small parameters $\sigma $ and $\gamma $ in any desirably wide range of
frequencies, $f_1=\gamma /\pi \sigma _\tau <f<f_2=1/\pi \sigma _\tau $, is
{\bf 1/f-like} \cite{K98,KM98,KM98VC,KM99C}, i.e.,
$$
S\left( f\right) =\bar I^2\sqrt{\frac 2\pi }\frac{\bar \tau \exp \left( -
\overline{\tau }^2/2\sigma _\tau ^2\right) }{\sigma _\tau f}.\eqno{(16)}
$$
Here $\sigma _\tau ^2=\sigma ^2/2\gamma $ is the variance of the interevent
time $\tau _k$.

\section{Origin of $1/f$ noise}

The origin for appearance of $1/f$ fluctuations in the model described in
Eqs. (14) and (15) is related with the relatively slow, Brownian,
fluctuations of the pulse interevent time. For this reason, the variance $
\sigma _\Delta ^2$ of the time difference $\Delta \left( k;q\right) $ for $
\left| q\right| \ll \gamma ^{-1}$ is a quadratic function of the time
difference and, consequently, of the difference $q$ of the pulse serial
numbers $k$ \cite{K98,KM98,KM98VC,KM99C}, i.e.,
$$
\sigma _\Delta ^2\left( k;q\right) =\sigma _\tau ^2\left( k\right) q^2.
\eqno{(17)}
$$

Substitution of Eqs. (10) and (17) into Eq. (8) yields $1/f$ spectrum (16).

\subsection{Generalization}

For slowly fluctuating interevent time, the time difference $\Delta \left(
k;q\right) $ may be expressed as \cite{K98,KM98,KM98VC,KM99C}
$$
\Delta \left( q\right) =\sum\limits_{l=k+1}^{k+q}\tau _l\simeq q\tau
\eqno{(18)}
$$
where $\tau =\left( t_{k+q}-t_k\right) /q$ is the average interevent time in
the time interval $\left( t_k,t_{k+q}\right) $, a slowly fluctuating
function of the arguments $k$ and $q$. In such an approach, the power
spectrum according to Eq. (6) is
$$
S\left( f\right) =2\bar Ia\sum\limits_q\left\langle e^{i2\pi fq\tau
}\right\rangle \eqno{(19)}
$$
where
$$
\left\langle e^{i2\pi fq\tau }\right\rangle \equiv \int\limits_{-\infty
}^{+\infty }e^{i2\pi fq\tau }\psi _\tau \left( \tau \right) {\rm d}\tau
=\chi _\tau \left( 2\pi fq\right)
$$
is the characteristic function of the distribution density $\psi _\tau (\tau
)$ of the interevent time $\tau $. Therefore, the power spectrum according
to Eq. (19) may be expressed as
$$
S\left( f\right) \simeq 2\overline{I}^2\overline{\tau }\Psi _\tau \left(
0\right) /f.\eqno{(20)}
$$
Here the property $\int_{-\infty }^{+\infty }\chi _\tau \left( x\right) {\rm
d}x=2\pi \Psi _\tau \left( 0\right) $ of the characteristic function has
been used.

\subsection{Correlation function of $1/f$ noise}

The correlation function of $1/f$ noise in the approximation (18) may be
calculated according to Eq. (9), i.e.,
$$
\begin{array}{c}
\Phi \left( s\right) =\bar Ia\sum\limits_q\int\limits_{-\infty }^{+\infty
}\psi _\tau (\tau )\delta \left( q\tau -s\right)
{\rm d}\tau \\ =\bar Ia\delta \left( s\right) +\bar Ia\sum\limits_{q\neq
0}\psi _\tau (\frac sq)\frac 1{\left| q\right| }.
\end{array}
\eqno{(21)}
$$
For the Gaussian distribution of the interevent time $\tau $
$$
\psi _\tau \left( \tau \right) =\frac 1{\sqrt{2\pi }\sigma _\tau }\exp
\left( -\frac{\left( \tau -\overline{\tau }\right) ^2}{2\sigma _\tau ^2}
\right)
$$
the correlation function (21) reads as
$$
\Phi \left( s\right) =\frac{\bar Ia}{\sqrt{2\pi }\sigma _\tau }\sum_q{\rm e}
^{-\frac{\left( s-q\overline{\tau }\right) ^2}{2\sigma _\tau ^2q^2}}\frac
1{\left| q\right| }.\eqno{(22)}
$$

It should be noted that the deviation of the variance $\sigma _\Delta ^2$
for large $q$ from the quadratic dependence (17) and approach to the linear
function $\sigma _\Delta ^2=2D_{t_k}\left| q\right| $ ensures the
convergence of sums (21) and (22) and, consequently, results in the
Lorentzian power spectrum (11) at $f\rightarrow 0$ \cite
{K98,KM98,KM98VC,KM99C}. Here $D_{t_k}$ is the ''diffusion'' coefficient of
the pulse occurrence time $t_k$, related with the variance $\sigma _{t_k}^2$
of the pulse occurrence time as $\sigma _{t_k}^2=2D_{t_k}k$. For the model
(14)--(15) $D_{t_k}=\sigma ^2/2\gamma ^2$.

The power spectra calculated according to Eq. (5) with the correlation
functions (21) and (22) are expressed as Eq. (20) and Eq. (16), respectively.

\section{Conclusions}

From the above analysis we can conclude that {\it the intrinsic origin of }$
1/f${\it \ noise is a Brownian fluctuations of the interevent time of the
signal pulses}, similarly to the Brownian fluctuations of the signal
amplitude resulting in $1/f^2$ noise. The random walk of the interevent time
in the time axis is a property of the randomly perturbed or complex systems
with the elements of selforganization.

\section*{ Acknowledgments}

The author acknowledges useful discussions with Prof. A. Bastys, Prof. R.
Katilius, Prof. A. Matulionis and Dr. T. Me\v skauskas.

\nocite{*} \bibliographystyle{IEEE}

\end{document}